\documentclass{iauc}
\usepackage{graphicx}

\title[The colours of Virgo dEs as seen by SDSS] 
{The colours of Virgo dEs as seen by SDSS}

\author[T.\ Lisker, E.\,K.\ Grebel, B.\ Binggeli]   
{Thorsten Lisker, Eva K.\ Grebel, \and Bruno Binggeli}

\affiliation{Astronomical Institute, University of Basel, Switzerland;
  email: tlisker@astro.unibas.ch}

\pubyear{2005}
\volume{198} 
\pagerange{xxx--xxx}
\date{xxx}
\jname{Near-Field Cosmology with Dwarf Elliptical Galaxies}
\editors{H.\ Jerjen and B.\ Binggeli, eds.}
\begin{document}

\maketitle

\begin{abstract}
A colour analysis of a carefully selected sample of over 200 Virgo cluster dEs
and 8 giant ellipticals (Es) in the SDSS yields the following
results: a) the nucleated dEs (dENs) follow a tight colour-magnitude
relation (CMR) which does not broaden significantly at fainter ($M_r
\ge -15$) magnitudes, b) the CMR of dENs smoothly transitions to 
the CMR of Es, but the latter changes its slope at about $M_r \approx
-20$, c) the brighter dENs are consistent with ages of $t \ge
5\,\rm{Gyr}$ and a decrease of metallicity towards fainter magnitudes, but a
possible additional luminosity-age relation is not ruled out, d) it is
crucial to treat dS0s separately from dEs since their different colour
properties would otherwise bias the comparison of nucleated and
non-nucleated dEs (dEnoNs), e) the dEnoNs show a weak trend towards
bluer colours than dENs.

\end{abstract}

\firstsection 
\section{Introduction and sample selection}

Numerous studies of dwarf ellipticals (dEs) have put the homogeneity of
this class of galaxy in question. In particular, the two main
subclasses, nucleated and non-nucleated dEs (dENs and dEnoNs), might
have a different history: dENs are more strongly
clustered, are brighter, and have been claimed to show redder colours
(Binggeli \& Cameron 1991; Rakos et al.\ 2004). To explore the
differences and similarities of these galaxies in greater detail, we
study here the fundamental relation of dEs --
the colour-magnitude relation (CMR) -- separately for dENs and
dEnoNs. This requires a large and homogeneous dE sample 
which ideally should also enable us to address the issue of a possible
transition from dwarf to giant ellipticals (Es). Contrary to the earlier belief,
it has been shown recently that the structural parameters of dEs and Es
exhibit a smooth transition (Graham \& Guzman 2003), raising the
question whether this is also the case for their colour properties.

The SDSS Data Release 3 provides
reduced and calibrated images in
the u, g, r, i, and z band as well as photometric measurements for
Virgo cluster galaxies. However, low surface
brightness objects like the dEs can be 'shredded',
i.e.\ parts of them are assigned to multiple separate source detections
by the detection algorithm, resulting in unreliable photometry. A
comparison with the work of Gavazzi et al.\ (2005) showed that
we must restrict ourselves to dEs brighter than $m_r \lesssim 18$
that are already listed in the Virgo Cluster Catalog (VCC; Binggeli et
al.\ 1985) and carefully exclude misaligned or shredded
source detections visually.
Moreover, the SDSS photometric pipeline
  overestimates the local sky flux due to the large apparent
  object sizes, for which we apply an appropriate correction. Our final
sample comprises 220 dEs+dS0s; the sample of 8
giant ellipticals was selected in the same manner. Colours for the dwarfs were derived
from aperture photometry on the SDSS images,
excluding the inner $r \le 2''$ to avoid nucleus light entering the
aperture. Since this does not require any photometric
measurements from the SDSS pipeline, our sample is extended to 349
dEs+dS0s in diagrams where only colours are required. Errors were estimated from the 
signal-to-noise ratio and the uncertainty of the
sky flux correction.
Galactic extinction corrections are provided by the SDSS database.

\section{Results}

The classification of several dwarfs as dS0 in the VCC
had been based on the two-dimensional appearance of the galaxies rather
than on their azimuthally averaged profiles -- the latter are
indistinguishable from those of the dEs (Binggeli \& Cameron
1991). Consequently, the existence of 
a separate class of dS0 galaxies has been questioned: Ryden et al.\
(1999)
conclude that there are no compelling arguments for it, and e.g.\
Barazza et al.\ (2003) subsume dEs and dS0s into one class. We
demonstrate the danger of this approach in Fig.\ 1, where we show the
CMR separately for dENs and dEnoNs, treating dS0s as dEs in the middle panel
but separating them in the right panel. From the middle panel one would
conclude that the CMR of dEnoNs shows a steeper slope than
the one of dENs, whereas the right panel reveals only a slight tendency towards bluer
colours of dEnoNs compared to dENs, also given the paucity of bright
dEnoNs. Here and in the following figures colour is measured within
the half-light radius; the left panel of Fig.\ 1 confirms that no
significant difference appears when using fixed apertures
instead.
The CMR of dENs is tight and does not broaden
towards fainter magnitudes, as opposed to the findings of previous
studies of dEs (e.g.\ Conselice et al.\ 2003). Note that the increase
in colour errors by more than a factor of two would require any
significant broadening to be larger than that, which is not the
case.

We extend the CMR to faint Es in Fig.\ 2, illustrating that there is a smooth transition from dEs
to Es which is mainly populated by dENs. The left 
panel shows the positions of various types 
of galaxy and demonstrates that compact faint 'M32-like' ellipticals
neither belong to the E nor to the dE class but define their own
colour-magnitude region instead. In the middle panel we show that the Es
basically follow the linear extension of the CMR of dENs as derived
from a least-squares fit. However, a study of nearby Es in the
SDSS by Chang et al.\ (2005) finds a steeper
slope which is well constrained for magnitudes $M_r \lesssim -20$ and
indicates a slope change at about this magnitude. In the right
panel we focus on the
CMR of dEnoNs in order to see whether or not they follow the relation
defined by dENs and Es. Unfortunately the small number of bright
dEnoNs prevents any definite conclusion -- despite a weak tendency
towards bluer colours the dEnoNs are not inconsistent with the slope defined
by the dENs.

A rough age and metallicity distribution of our sample can be obtained
through a multicolour comparison of the observations with population
synthesis models from Bruzual \& Charlot (2003), as shown in Fig.\
3. Since the range of values for $i-z$ is rather small, we select only
objects brighter than $m_r \le 16$ to avoid large errors that would
dilute the comparison.
Brighter objects tend to be redder because of the CMR (left
panel of Fig.\ 3), therefore
the colours can only be explained with a decrease in metallicity
towards fainter magnitudes (right panel), consistent with previous
studies.  Young ages ($t \lesssim 2\,\rm{Gyr}$) are clearly ruled out,
and for the dENs the data is consistent with ages $t \gtrsim 
5\,\rm{Gyr}$. However, we cannot exclude a possible relation of age and
luminosity \emph{on top} of the dominant luminosity-metallicity
relation; moreover, the trend of dEnoNs showing bluer colours appears
to be strongest in $u-g$, possibly indicating younger ages than for the
dENs.

\section{Conclusions}\label{sec:concl}
Our large sample of Virgo cluster dEs+dS0s from the SDSS shows that dENs
follow a tight CMR down to $M_r \approx -13$ without any significant broadening in
colour.
The faint Es follow
the CMR defined by dENs, but a change in slope occurs at $M_r \approx
-20$. After excluding dS0
galaxies, only a weak trend towards bluer colours of dEnoNs compared to
dENs is found. The colours of the brighter dEs rule out young ages ($t \lesssim
2\,\rm{Gyr}$) and the dENs are consistent with ages $t \gtrsim
5\,\rm{Gyr}$.

\begin{acknowledgments}
    We thank the SDSS collaboration for the wealth of DR3
    data. The SDSS Web site is http://www.sdss.org/.
\end{acknowledgments}

\begin{discussion}

\discuss{Ferguson}{Quite a few galaxies in your colour-colour diagram
  were inconsistent with any of the models. Are you confident that
  there are not systematic uncertainties in the models or data? How far
  do you think you could shift the model colours, e.g.\ using different
  isochrones?}
\discuss{Lisker}{I was using Padova 2000 isochrones against the
  recommendation of Bruzual \& Charlot (2003), since the
  Padova 1994 isochrones were offset towards redder $i-z$ colours by a
  noticeable amount -- so indeed, we are facing a 'model error' in
  addition to the observational error.}
\discuss{Rakos}{Would you agree that the difference in the slope you
  have shown between E and dE galaxies is introduced by different
  metallicity?}
\discuss{Lisker}{Since both the shallower slope of the CMR at $M_r \lesssim
  -20$ and the steeper slope at $M_r \gtrsim
  -20$ are luminosity-metallicity relations, I am not sure whether
  the \emph{change} in the slope can be caused by metallicity itself.}
\discuss{Conselice}{If you were to include all of the fainter dEs from
  the VCC that are not included in your sample, you would probably see
  a larger scatter in the CMR at faint magnitudes.}
\discuss{Lisker}{We cannot test this so far, since the SDSS photometry
  catalogs are not useful for the faintest dEs, mainly due to
  galaxy shredding. However, we already go down to $M_r \approx -13$,
  clearly fainter than where a significant increase in the colour
  spread has been claimed to appear.}
\end{discussion}

\begin{figure}
\centering
\resizebox{12.5cm}{!}{\rotatebox{0}{\includegraphics{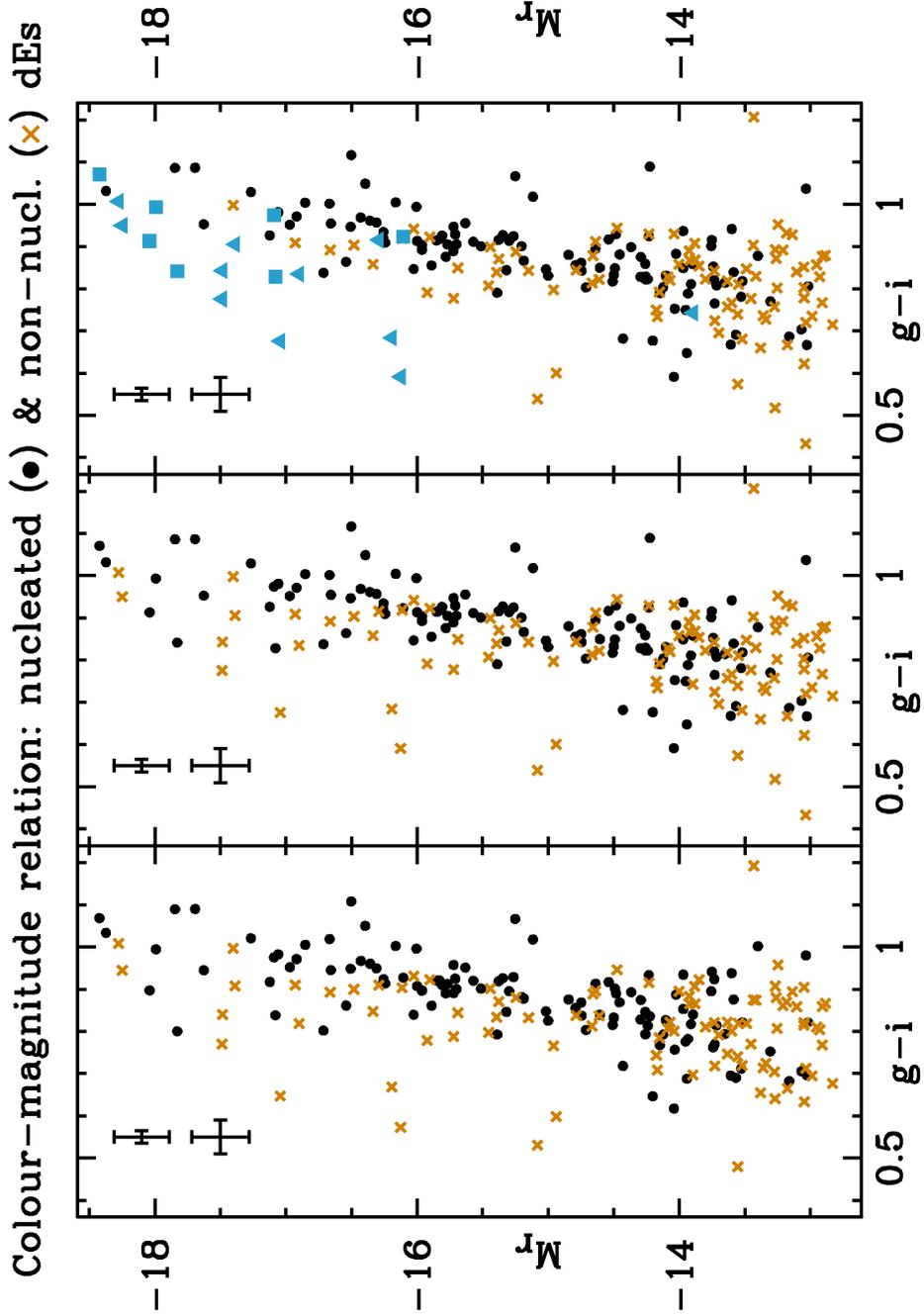}}}
  \caption{CMR for dENs (filled circles), dEnoNs
    (crosses), nucleated dS0s (squares) and non-nucleated dS0s
    (triangles). Upper and lower error bars indicate mean magnitude
    uncertainties for $M_r\lesssim -15.5$ and $M_r\gtrsim -15.5$,
    respectively.
    A distance modulus of
    $m-M=31$ is adopted. \emph{Left:} Absolute r-band magnitude versus
    fixed aperture ($r \approx 8''$) $g-i$ colour, including dS0s into the dE
    class. \emph{Middle:} CMR with $g-i$ colour measured from half-light
    radius apertures, including dS0s into the dE class. \emph{Right:}
    CMR with $g-i$ colour measured from half-light radius apertures, now
    showing dS0s separately.}
\label{fig1}
\end{figure}

\begin{figure}
\centering
\resizebox{12.5cm}{!}{\rotatebox{0}{\includegraphics{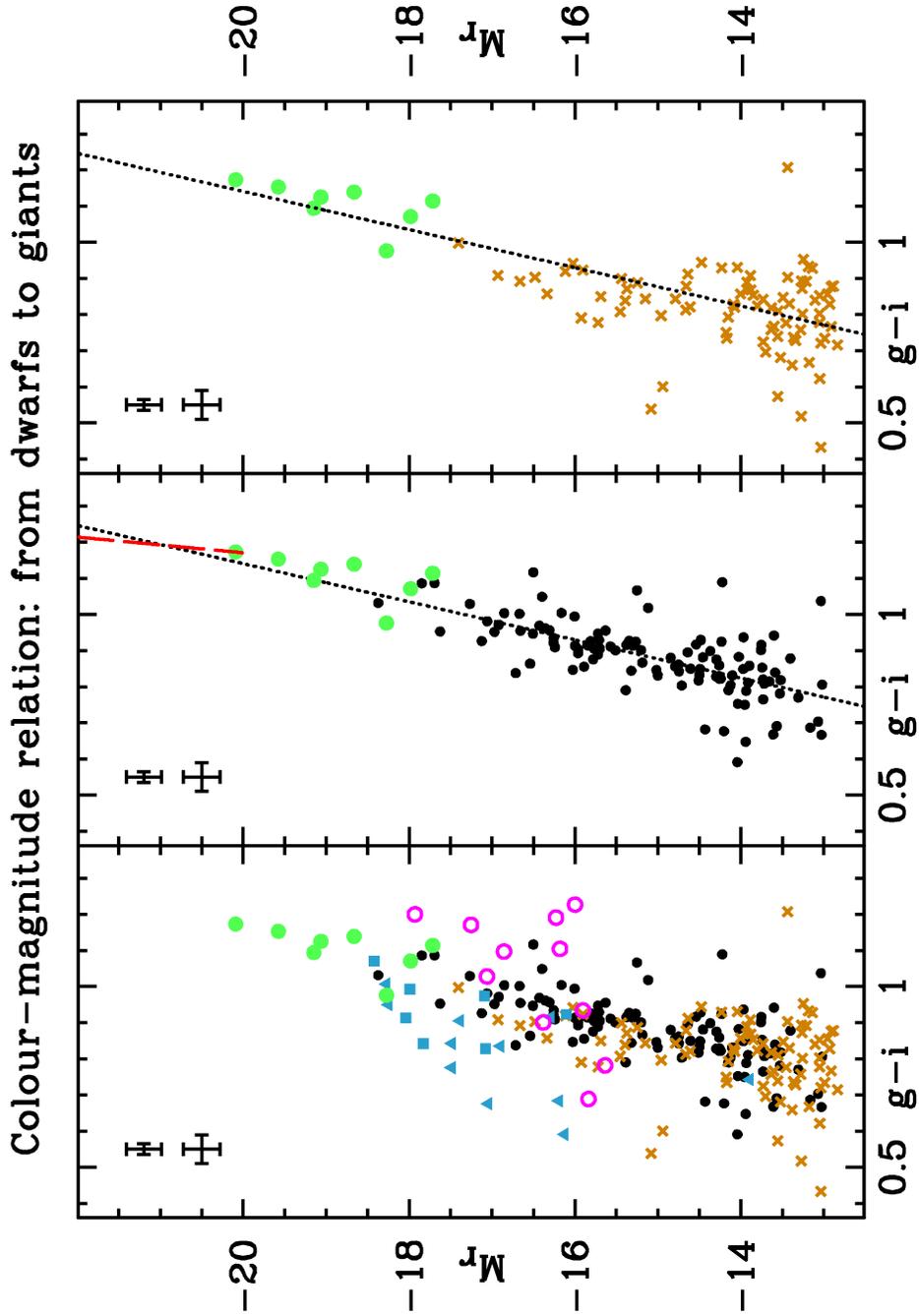}}}
  \caption{\emph{Left:} CMR for dENs (filled circles), dEnoNs
    (crosses), nucleated dS0s (squares), non-nucleated dS0s
    (triangles), compact faint (M32-like) ellipticals (open circles),
    and Es (large filled circles). \emph{Middle:} CMR for dENs and Es
    only. The dotted line shows a least-squares fit to the dENs, the
    dashed line is the slope of the CMR of brighter giant ellipticals
    from Chang et al.\ 2005. \emph{Right:} CMR for dEnoNs and Es
    only. The dotted line is the same as in the middle panel.}
\label{fig2}
\end{figure}

\begin{figure}
\centering
\resizebox{12.5cm}{!}{\rotatebox{0}{\includegraphics{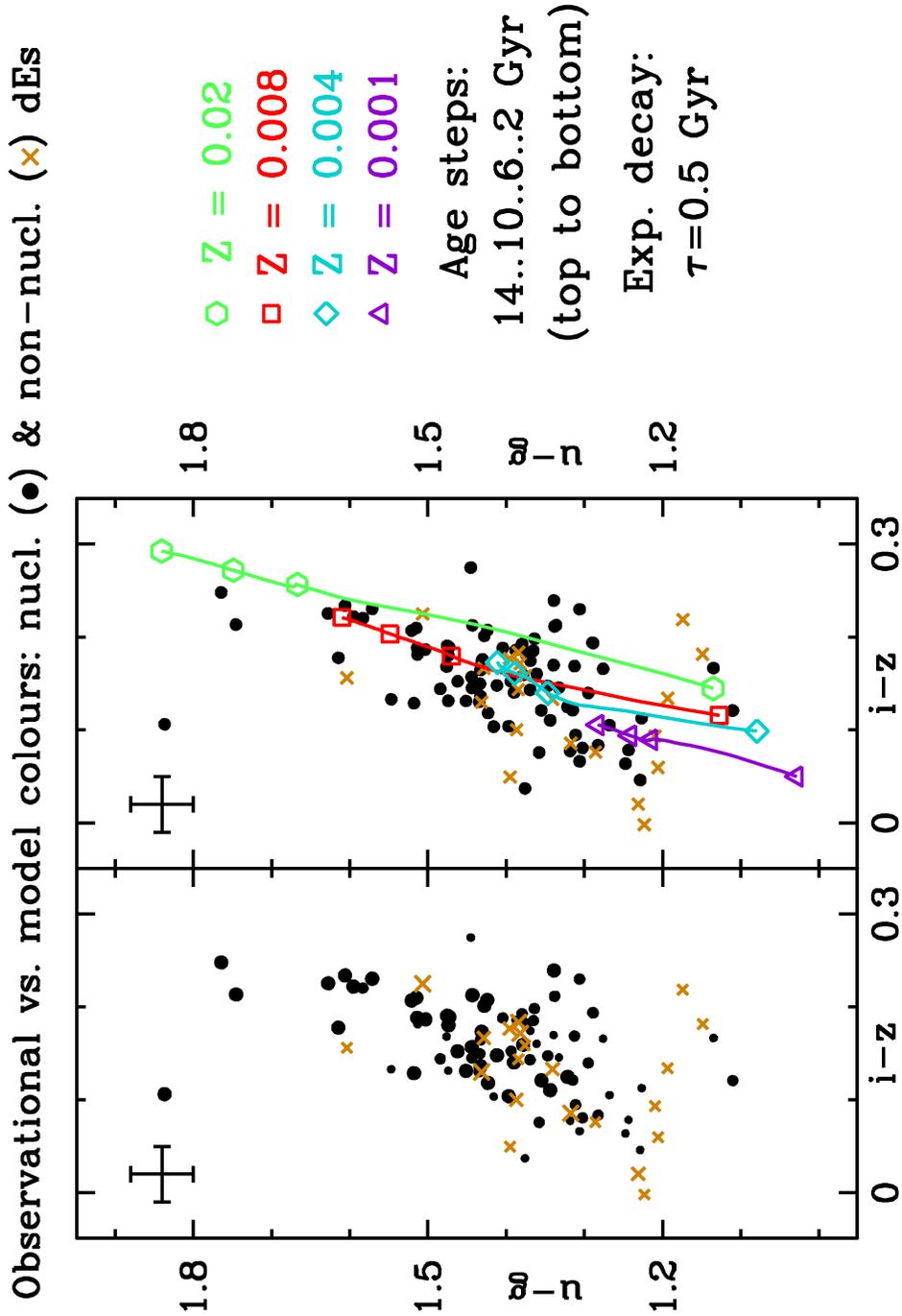}}}
  \caption{Colour-colour plot for dENs (filled circles) and dEnoNs
    (crosses).
Only objects brighter than $m_r \le 16$ were
    selected to ensure small errors in the colour. \emph{Left:} The
    effect of the CMR is illustrated by the symbol sizes: larger for
    brighter objects.
\emph{Right:} Comparison of observed and
    model colours. Each track from Bruzual \& Charlot (2003) shows the
    colours of a single-burst stellar population with a decay time of
    $\tau = 0.5\,\rm{Gyr}$ and an age of 2 (bottom) to 14 (top) Gyr
    using Padova 2000 isochrones and a Chabrier IMF.}
\label{fig3}
\end{figure}

\end{document}